\documentclass[prd,twocolumn,showpacs,aps,epsbox]{revtex4}
\usepackage{epsfig}

\def\beq{\begin{equation}}
\def\eeq{\end{equation}}
\def\bea{\begin{eqnarray}}
\def\eea{\end{eqnarray}}
\def\nn{\nonumber\\}
\def\pa{\partial}
\def\ra{\rightarrow}
\def\bp{\mbox{\boldmath$\phi$}}
\def\tm{\tilde{m}}
\def\te{\tilde{e}}
\def\tq{\tilde{q}}
\def\tE{\tilde{E}}
\def\tQ{\tilde{Q}}

\begin{document}

\draft
OCU-PHYS-341, AP-GR-85
\title{Q-tubes and Q-crusts}
\author{Nobuyuki Sakai}
\email{nsakai@e.yamagata-u.ac.jp}
\affiliation{Department of Education, Yamagata University, Yamagata 990-8560, Japan}
\author{Hideki Ishihara}
\email{ishihara@sci.osaka-cu.ac.jp}
\author{Ken-ichi Nakao}
\email{knakao@sci.osaka-cu.ac.jp}
\affiliation{Department of Physics, Osaka City University, Osaka 558-8585, Japan}

\begin{abstract}
We explore equilibrium solutions of non-topological solitons in a general class of scalar field theories which include global U(1) symmetry.
We find new types of solutions, tube-shaped and crust-shaped objects, and investigate their stability.
Like Q-balls, the new solitons can exist in supersymmetric extensions of the Standard Model, which may responsible for baryon asymmetry and dark matter.
Therefore, observational signals of the new solitons would give us more informations on the early universe and supersymmetric theories.
\end{abstract}

\maketitle

\section{Introduction}

In a pioneering work by Friedberg, Lee and Sirlin in 1976 \cite{FLS76}, non-topological solitons were introduced in a model with a U(1)-symmetric complex scalar field coupled to a real scalar field.
In contrast with topological defects, they are stabilized by a global U(1) charge, and their energy density is localized in a finite space region without gauge fields.
In 1985 Coleman showed such solitons exist in a simpler model with an SO(2) (viz.\ U(1)) symmetric scalar field only, and called them Q-balls \cite{Col85}.

Q-balls have attracted much attention in particle cosmology since Kusenko pointed out that they can exist in all supersymmetric (SUSY) extensions of the Standard Model \cite{Kus97b-98}.
Specifically, Q-balls can be produced efficiently in the Affleck-Dine mechanism \cite{AD} and could be responsible for baryon asymmetry \cite{SUSY} and dark matter \cite{SUSY-DM}.
Q-balls can also influence the fate of neutron stars \cite{Kus98}.
Based on these motivations, stability of Q-balls has been intensively studied \cite{stability,PCS01,SS,TS}.

Observational signatures of SUSY Q-balls has been studied \cite{ex},
and their mass and flux were constrained by experimental data of the searches for magnetic monopoles and heavy cosmic rays \cite{AYNO00}.
Currently direct searches for neutral Q-balls and for electrically charged Q-balls are in progress in Super-Kamiokande II \cite{SKII} and in the SLIM Experiment \cite{SLIM}, respectively.
Furthermore, it has been shown that gravitational waves are emitted during Q-ball formation and could be detected by next-generation gravitational detectors \cite{GW},

In spite of increasing concern about non-topological solitons in SUSY, other equilibrium solutions has not been studied so much, while topological defects have several types according to the symmetry.
In this paper we address a fundamental question: are their any other non-topological solitons in different forms?
If new solitons exist in the theories which allow for Q-balls, their observational signals would give us new informations on SUSY Q-ball models.
Here we re-analyze scalar field theories which include global U(1) symmetry.

\section{Equilibrium solutions}

We begin with a review of Q-ball solutions.
Consider an SO(2)-symmetric scalar field $\bp=(\phi_1,\phi_2)$, whose action is given by
\beq\label{S}
{\cal S}=\int d^4x\left[
-\frac12\eta^{\mu\nu}\pa_{\mu}\bp\cdot\pa_{\nu}\bp-V(\phi) \right],
~~\phi\equiv\sqrt{\sum_{a=1}^2 \phi_a \phi_a}.
\eeq
Due to the symmetry there is a conserved charge,
\beq
Q\equiv\int d^3x\left(\phi_1{\pa\phi_2\over\pa t}-\phi_2{\pa\phi_1\over\pa t}\right).
\eeq
Assuming spherical symmetry and homogeneous phase rotation,
\beq\label{qball-phase}
\bp=\phi(r)(\cos\omega t,\sin\omega t),
\eeq
one has a field equation,
\beq\label{FEqball}
{d^2\phi\over dr^2}+\frac2r{d\phi\over dr}+\omega^2\phi={dV\over d\phi}.
\eeq
This is equivalent to the field equation for a single static scalar field with a potential 
$V_{\omega}=V-\omega^2\phi^2/2$.

Equilibrium solutions $\phi(r)$ with a boundary condition
\beq\label{BCqball}
{d\phi\over dr}(r=0)=0,~~~\phi(r\ra\infty)=0,
\eeq
exist if min$(V_{\omega}) <V_{\omega}(0)$ and $dV_{\omega}/d\phi(0) > 0$. This condition is rewritten as
\beq\label{Qexist}
{\rm min}\left[{2(V-V(0))\over\phi^2}\right]<\omega^2<m^2\equiv{d^2V\over d\phi^2}(0).
\eeq

\begin{figure}[htbp]
\psfig{file=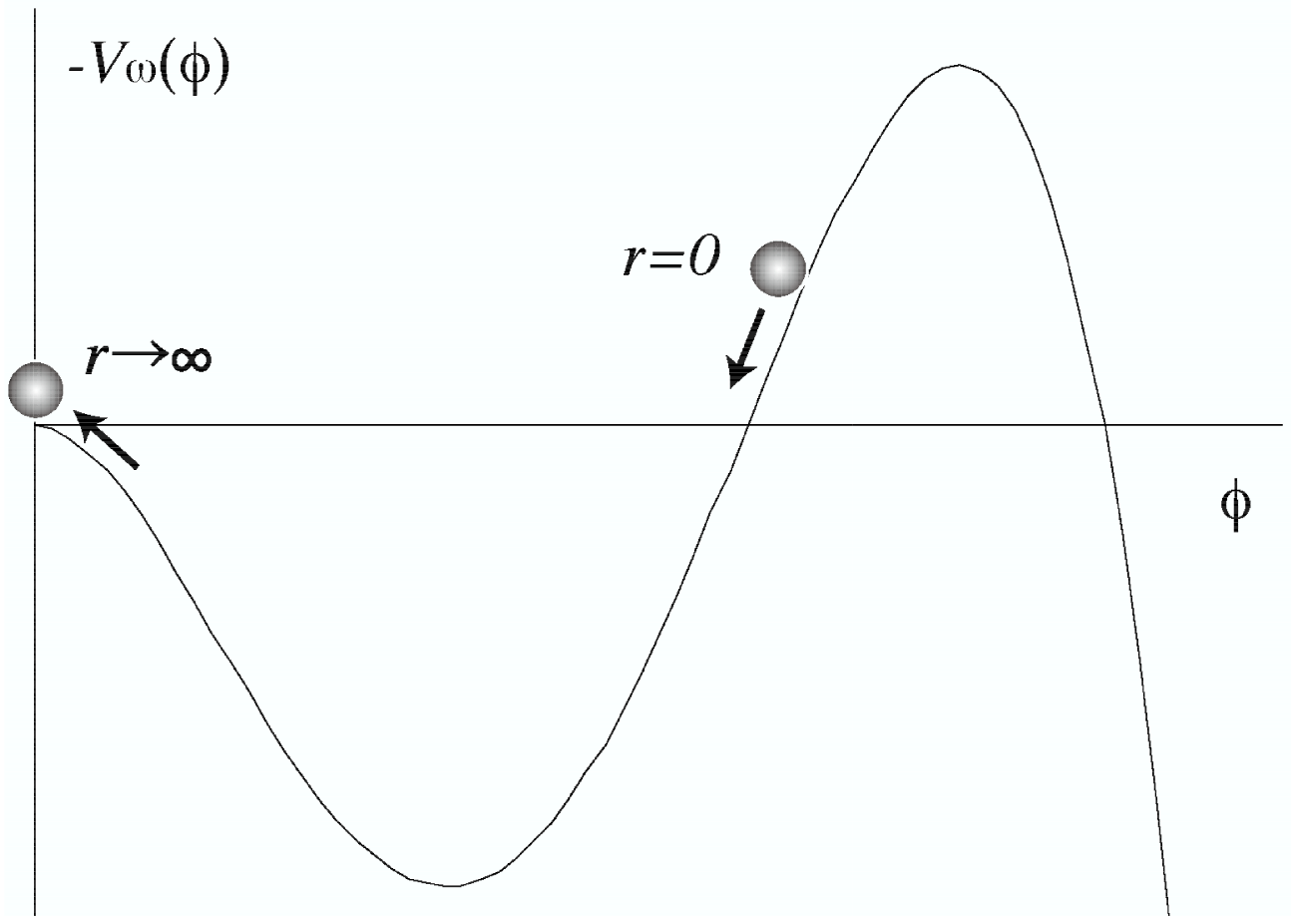,width=3in}
(a)
\psfig{file=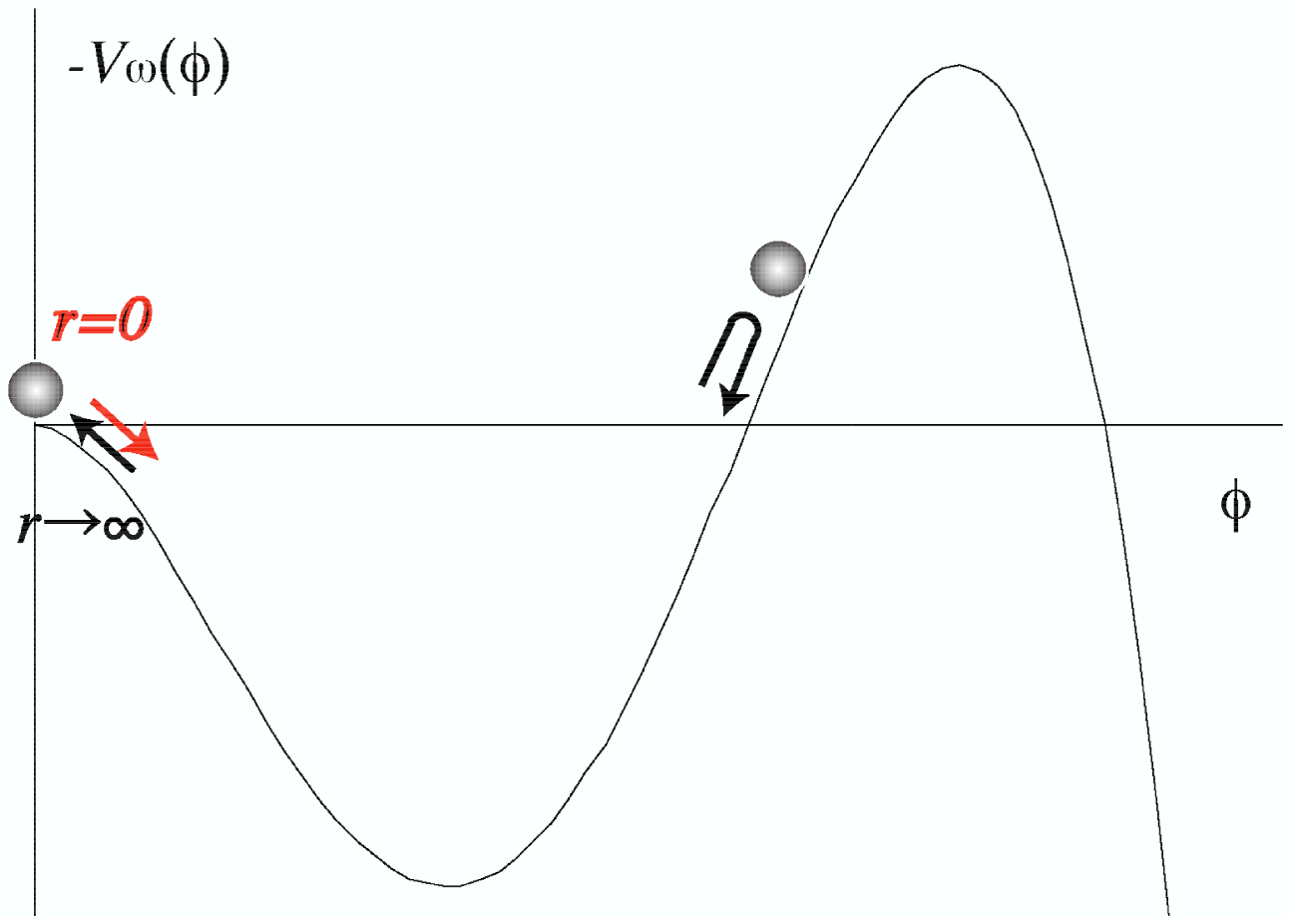,width=3in}
(b)
\caption{\label{Newton}
Interpretation of (a) Q-ball solutions and (b) new soliton solutions by analogy with a particle motion in Newtonian mechanics.}
\end{figure}

If one regards the radius $r$ as \lq time\rq\ and the scalar amplitude $\phi(r)$ as \lq the position of
a particle\rq, one can understand Q-ball solutions in words of Newtonian mechanics, as shown in Fig.\ \ref{Newton}(a).
Equation (\ref{FEqball}) describes a one-dimensional motion of a particle under the conserved force due to the potential $-V_{\omega}(\phi)$ and the \lq time\rq-dependent friction $-(2/r)d\phi/dr$.
If one chooses the \lq initial position\rq\ $\phi(0)$ appropriately, the static particle begins to roll down the potential slope, climbs up and approaches the origin over infinite time. 

To demonstrate numerical solutions later, we adopt  a simple model,
\beq\label{V3}
V=\frac12m^2\phi^2-\mu\phi^3+\lambda\phi^4,
~~{\rm with}~~
m^2,~\mu,~\lambda>0.
\eeq
and rescale the quantities as
\beq\label{rescale}
\tilde{x^{\mu}}\equiv{\mu\over\sqrt{\lambda}}x^{\mu},~
\tilde{\phi}\equiv{\lambda\over\mu}\phi,~
\tm\equiv{\sqrt{\lambda}\over\mu}m,~
\tilde{\omega}\equiv{\sqrt{\lambda}\over\mu}\omega.
\eeq
Then, the existing condition (\ref{Qexist}) becomes
\beq\label{QexistV3}
0<\epsilon^2<\frac12,~~~
\epsilon^2\equiv\tm^2-\tilde{\omega}^2.
\eeq

\vskip 3mm
{\it (1) Q-tubes.}
For the same SO(2) model, we suppose a string-like configuration,
\beq\label{qtube-phase}
\bp=\phi(R)(\cos(n\varphi+\omega t),\sin(n\varphi+\omega t)),
\eeq
where $n$ is nonnegative integer and $(R,\varphi,z)$ is the cylindrical coordinate system.
The field equation becomes
\beq\label{FEtube}
{d^2\phi\over dR^2}+\frac1R{d\phi\over dR}-{n^2\phi\over R^2}+\omega^2\phi={dV\over d\phi}.
\eeq

\begin{figure}[htbp]
\psfig{file=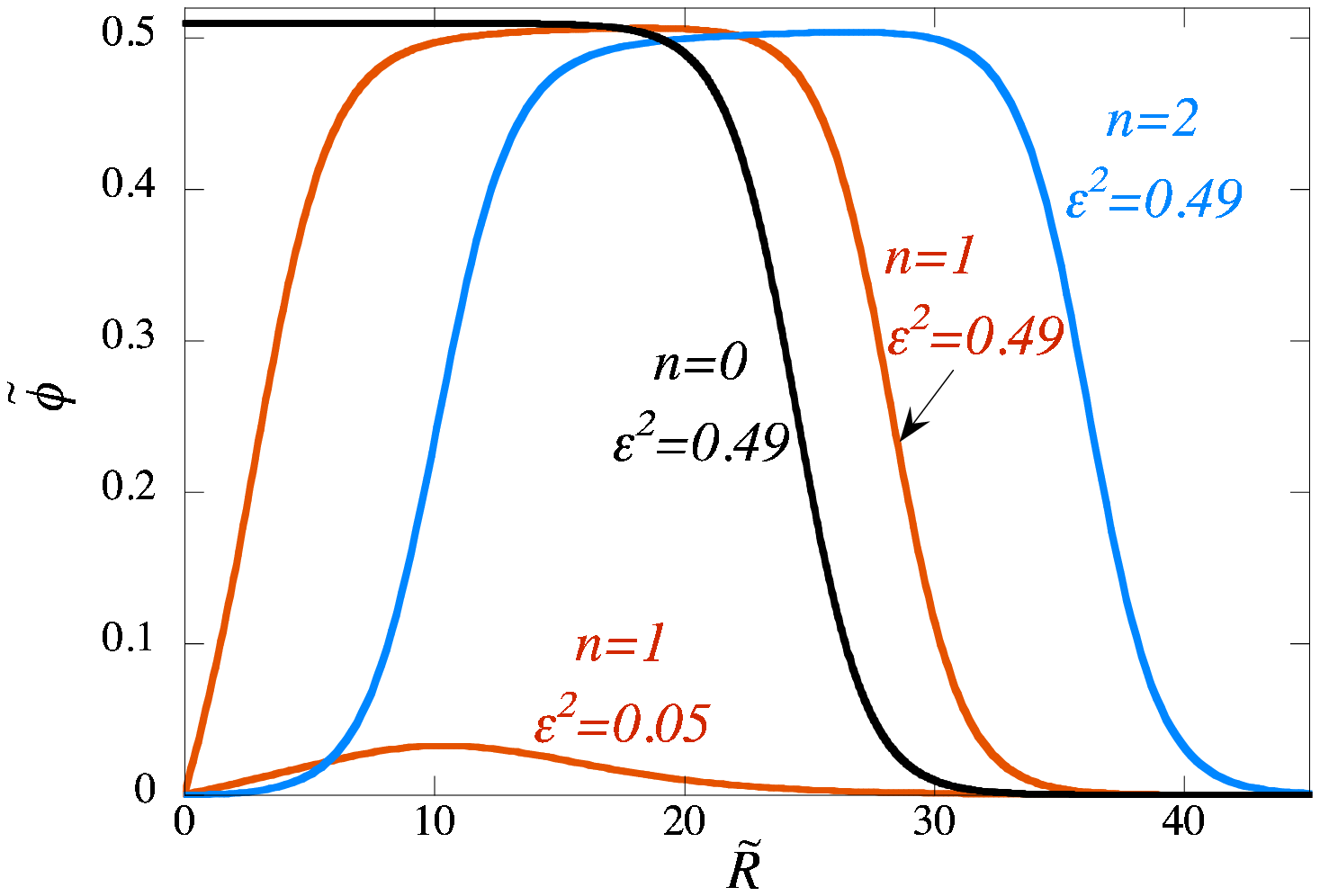,width=3in}
(a)
\psfig{file=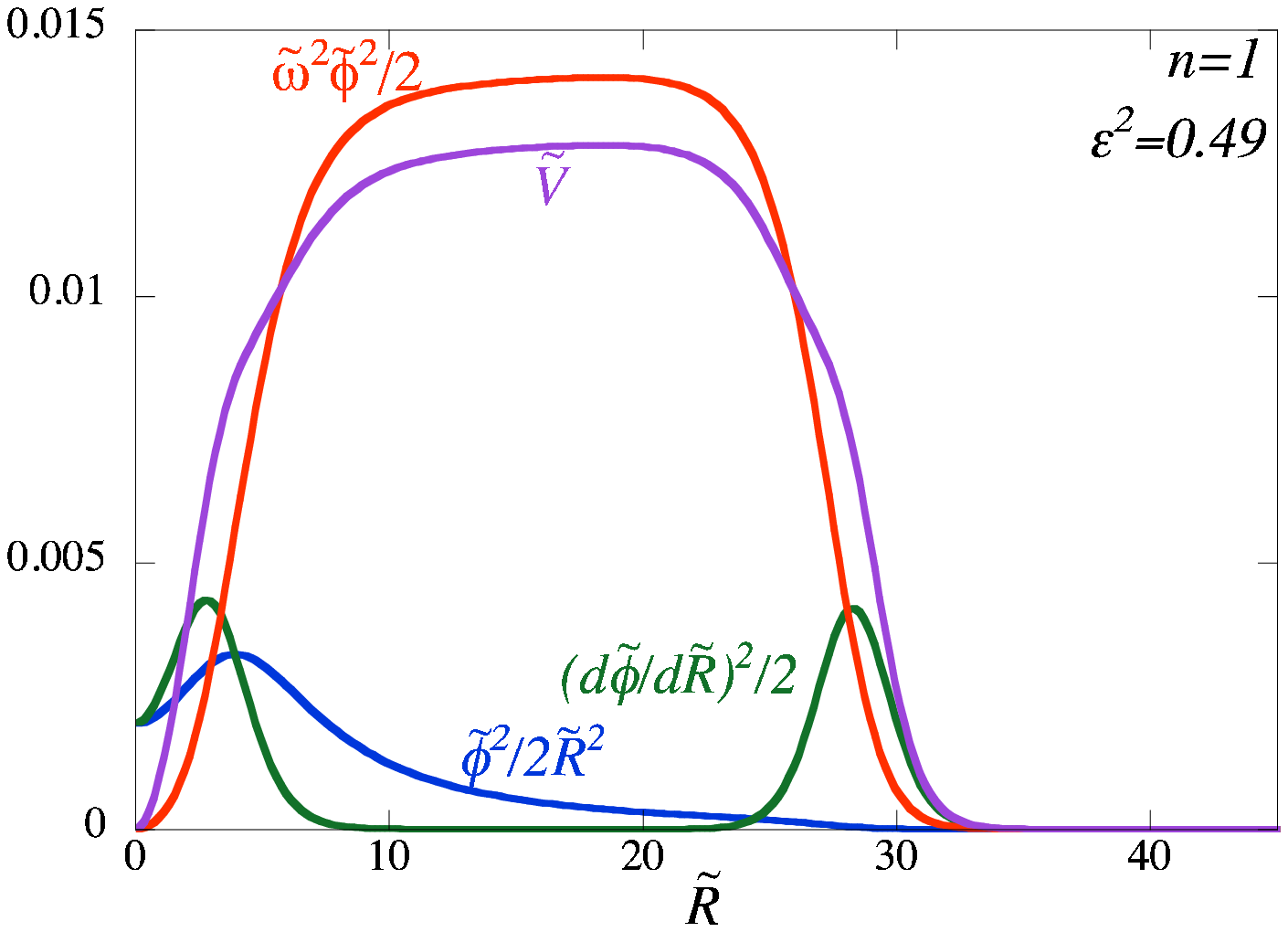,width=3in}
(b)
\psfig{file=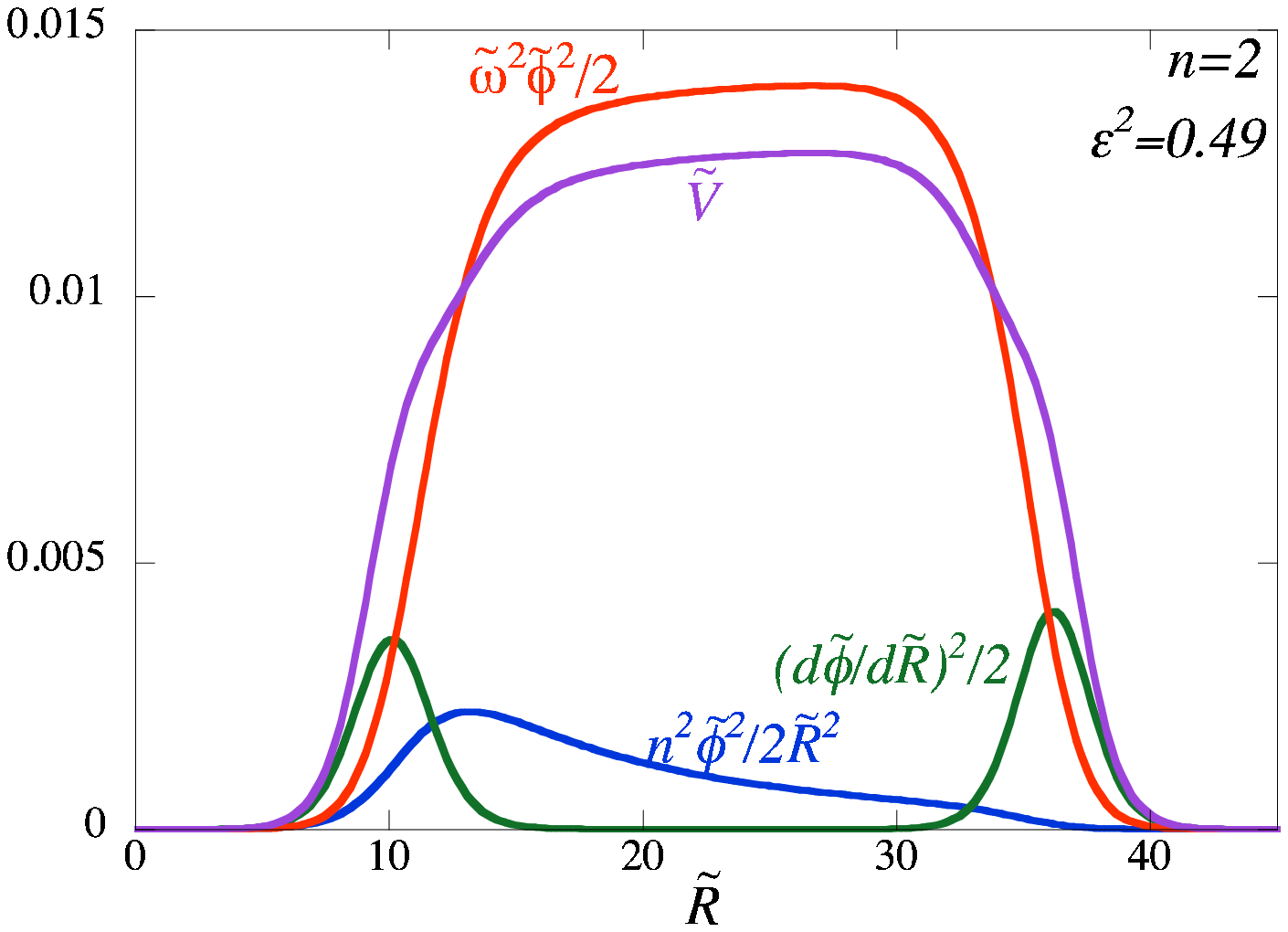,width=3in}
(c)
\caption{\label{qtube}
Examples of Q-tube solutions.
(a) shows $\phi(r)$ for four cases.
(b) and (c) show each term of the energy-momentum tensor for two of the solutions,
where we put $\tm^2=0.6$.}
\end{figure}

If $n=0$, the field equation is the same as (\ref{FEqball}) except for a numerical coefficient.
Therefore, Q-ball like solutions of $\phi(R)$ exist.
If $n\ge1$,  there is no regular solution which satisfies $\phi(0)\ne0$.
However, if we adopt a different boundary condition, 
\beq
\phi(R=0)=\phi(R\ra\infty)=0,
\eeq
there is a new type of regular solutions.
We introduce an auxiliary variable $\psi$ which is defined by
$\phi(R)=R^n\psi(R)$,
Then, Eq.(\ref{FEtube}) becomes
\beq
{d^2\psi\over dR^2}+{2n+1\over R}{d\psi\over dR}+\omega^2\psi
=R^{-n}{dV\over d\phi}\Big|_{\phi=R^n\psi}
\eeq
If we choose $\psi(0)$ appropriately, we obtain a solution $\psi(R)$ which is expressed in the Maclaurin series without odd powers in the neighborhood of $R=0$.
In terms of the original variable $\phi(R)$, the $n$th differential coefficient $\phi^{(n)}(0)=\psi(0)$ should be determined by the shooting method, while any lower derivative vanishes at $R=0$.
We plot some solutions in Fig.\ \ref{qtube}(a).

We can illustrate existence of the new solutions with $n\ge1$ by analogy with a particle motion in Newtonian mechanics, as shown in Fig.\ \ref{Newton}(b).
Equation (\ref{FEtube}) describes a one-dimensional motion of the particle under the conserved force due to the potential $-V_{\omega}(\phi)$ and two non-conserved forces, the friction $-(1/R)d\phi/dR$ and the repulsive force $n^2\phi^2/R^2$.
If $n=1$,  by choosing the  \lq initial velocity\rq\ $d\phi/dR(0)$ appropriately, the particle goes down and up the slope, and at some point $\phi=\phi_{{\rm max}}$ it turns back and approaches the origin over infinite time.
If $n\ge2$, $d\phi/dR(0)$ vanishes; instead, the $n$th derivative $\phi^{(n)}(0)$ gently pushes the particle at $\phi=0$. Therefore, with the appropriate choice of $\phi^{(n)}(0)$, the particle moves along a similar trajectory to that of $n=1$.
This argument also indicates that the existence condition of the new soliton solutions are the same as that of Q-balls, (\ref{Qexist}) or (\ref{QexistV3}).
Solutions with the same behavior as the $n=1$ solutions were obtained by Kim {\it et al}.\cite{Kim}, who studied the SO(3)-symmetric scalar field without Q-charge.

Nonzero components of the energy-momentum tensor are given by
\bea
-T^t_t&=&\frac12\left({d\phi\over dR}\right)^2+\frac{n^2\phi^2}{2R^2}+{\omega^2\over2}\phi^2+V,\nn
T^R_R&=&\frac12\left({d\phi\over dR}\right)^2-\frac{n^2\phi^2}{2R^2}+{\omega^2\over2}\phi^2-V,\nn
T^{\varphi}_{\varphi}&=&-\frac12\left({d\phi\over dR}\right)^2+\frac{n^2\phi^2}{2R^2}+{\omega^2\over2}\phi^2-V,\nn
T^z_z&=&-\frac12\left({d\phi\over dR}\right)^2-\frac{n^2\phi^2}{2R^2}+{\omega^2\over2}\phi^2-V,\nn
T^t_{\varphi}&=&-n\omega\phi^2
\eea
The nonzero component $T^t_{\varphi}$ indicates that the solutions possess angular-momentum.
Each term of the energy-momentum tensor is presented in Fig.\ \ref{qtube}(b)(c).
Although the present analysis does not include gravity,
we can estimate gravitational effects in the weak field approximation as follows.
If we define the gravitational potential $\Phi$ as $\Phi\equiv g_{tt}+1$, and take it into account up to its first order,
the Einstein equations yeild the extended Poisson equation,
\beq\label{Poisson}
\partial^{\mu}\partial_{\mu}\Phi=4\pi G(-T^t_t+T^i_i)=8\pi G(\omega^2\phi^2-V),
\eeq
where $T^i_i$ is the trace of the spatial components.
Except for $n=0$ cases, the gravitational source in (\ref{Poisson}) vanishes in the center in the $x$-$y$ plane.
Furthermore, this field configuration has planar symmetry in the $z$ direction.
Therefore, we collectively call the new solitons Q-tubes.

Because SO(2) is a subgroup of SO($N\ge3$) or SU($N\ge2$), Q-tubes with $n=0$ as well as Q-balls can appear in any SO($N$) or SU($N$) theory.
By contrast, we suspect that Q-tubes with $n\ge1$ can appear only in U(1) theory, since their string-like configuration is topologically unstable in theories with $SO(N\ge3)$ or $SU(N\ge2)$.

\begin{figure}[htbp]
\psfig{file=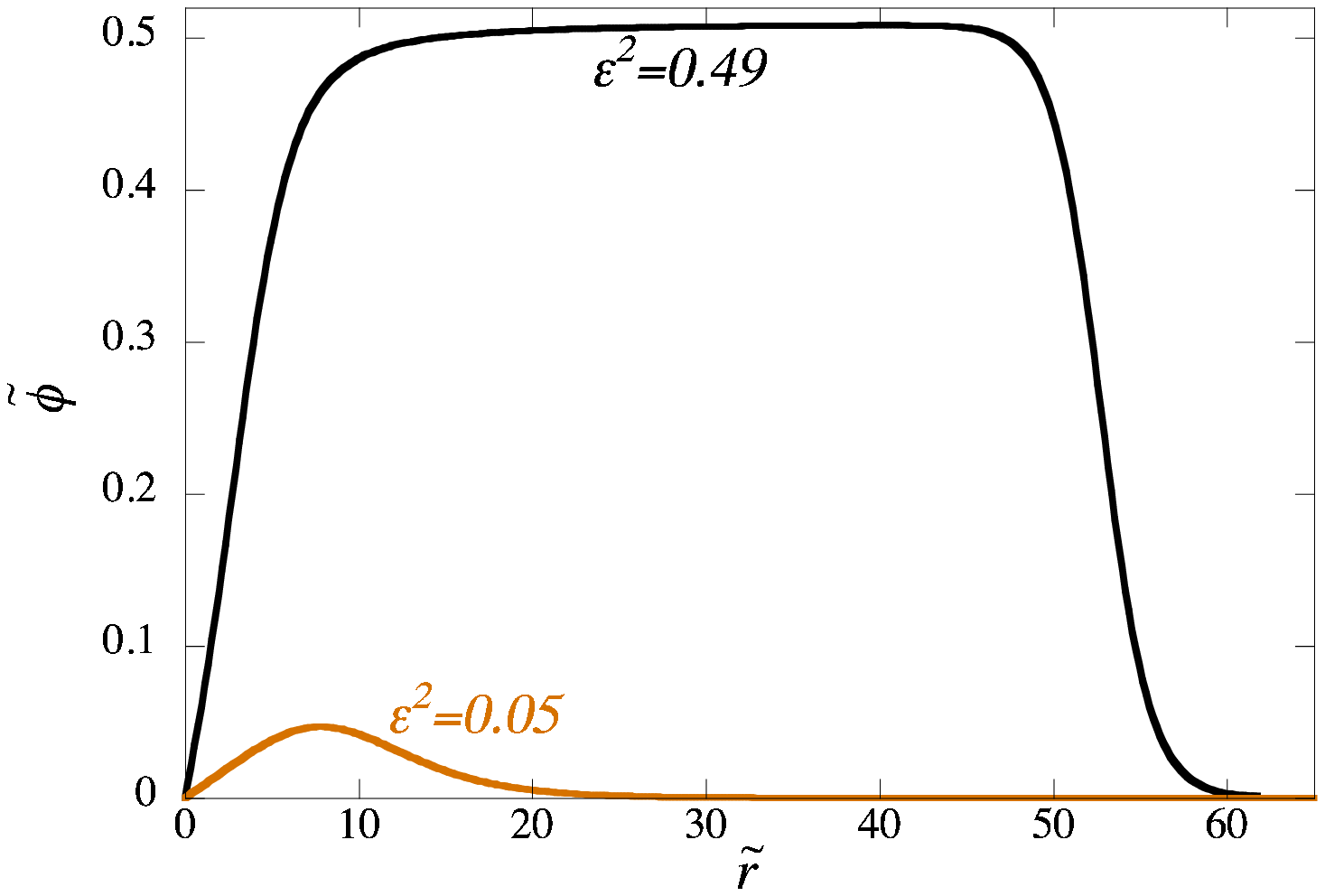,width=3in}
(a)
\psfig{file=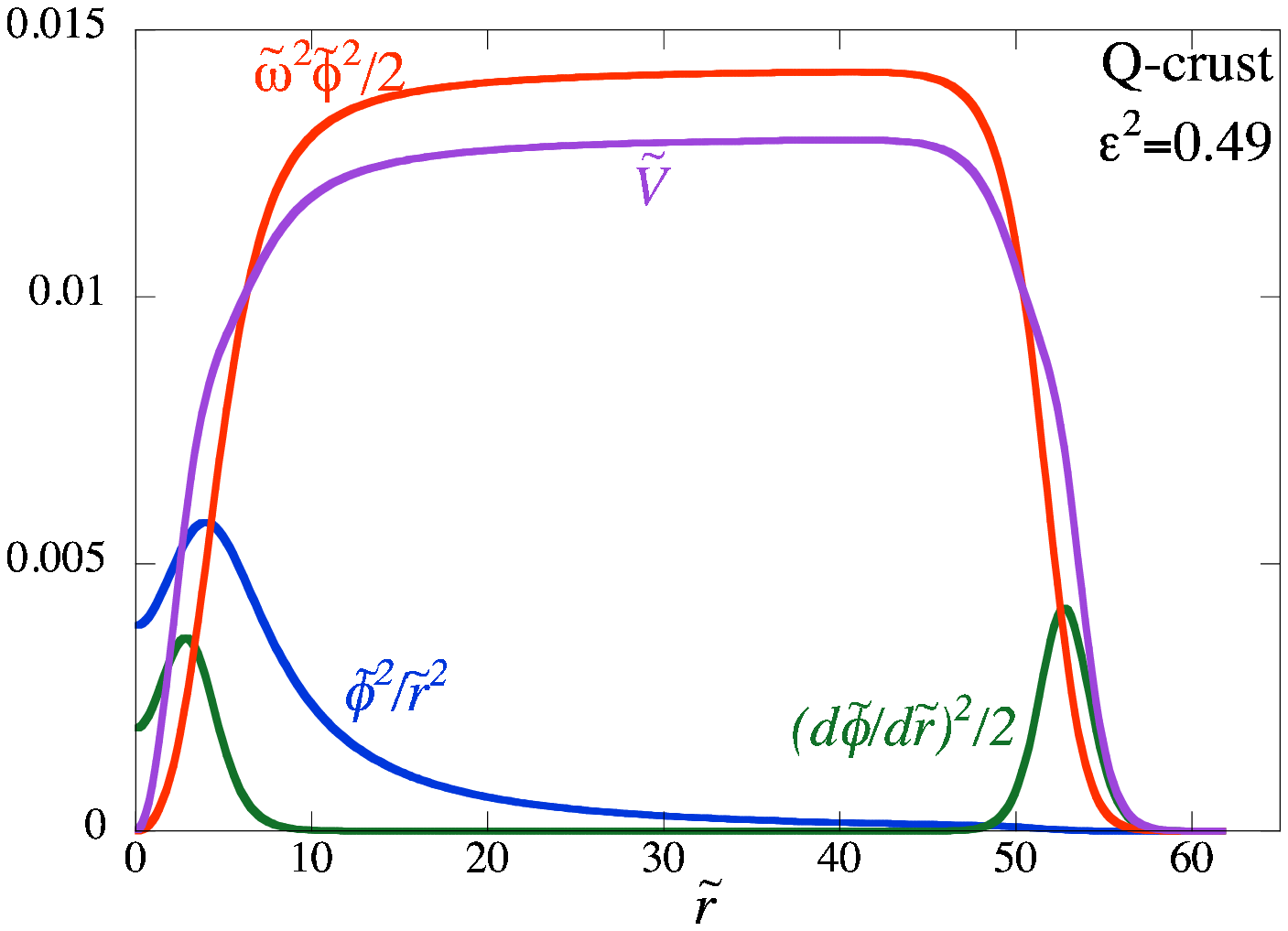,width=3in}
(b)
\caption{\label{qcrust}
Examples of Q-crust solutions.
(a) shows $\phi(r)$ for two cases.
(b) shows each term of the energy-momentum tensor for the solution with $\epsilon^2=0.49$.}
\end{figure}

\vskip 3mm
{\it (2) Q-crusts.}
Next, we consider an SO(3)$\times$U(1)-symmetric scalar field $\bp=e^{i\chi}(\phi_1,\phi_2,\phi_3)$, whose action is given by
\beq{\cal S}=\int d^4x\left[
-\frac12\eta^{\mu\nu}{\pa_{\mu}}\bp^{\ast}\cdot{\pa_{\nu}}\bp
-V(\phi) \right],~~~
\phi \equiv\sqrt{\sum_{a=1}^3\phi_a\phi_a}
\eeq
Assuming
\beq\label{qcrust-phase}
\bp=e^{i\omega t}\phi(r)(\cos\varphi\sin\theta,\sin\varphi\sin\theta,\cos\theta),
\eeq
we obtain the field equation,
\beq\label{FEcrust}
{d^2\phi\over dr^2}+\frac2r{d\phi\over dr}-{2\phi\over r^2}+\omega^2\phi={dV\over d\phi}.
\eeq
If we adopt a boundary condition,
\beq\label{BCcrust}
\phi(r=0)=\phi(r\ra\infty)=0,
\eeq
we find a regular solution $\phi(r)$, like a Q-tube solution with $n=1$.
We plot some solutions in Fig.\ \ref{qcrust}(a).

Nonzero components of the energy-momentum tensor are given by
\bea
-T^t_t&=&\frac12\left({d\phi\over dr}\right)^2+\frac{\phi^2}{r^2}+{\omega^2\over2}\phi^2+V,\nn
T^r_r&=&\frac12\left({d\phi\over dr}\right)^2-\frac{\phi^2}{r^2}+{\omega^2\over2}\phi^2-V,\nn
T^{\theta}_{\theta}=T^{\varphi}_{\varphi}&=&-\frac12\left({d\phi\over dr}\right)^2+{\omega^2\over2}\phi^2-V.
\eea
In this case $T^t_{\varphi}$ vanishes, which means that the solutions possess no angular-momentum.
Each term of the energy-momentum tensor is also shown by Fig.\ \ref{qcrust}(b).
The extended Poisson equation is given by
\beq
\partial^{\mu}\partial_{\mu}\Phi=4\pi G(-T^t_t+T^i_i)=8\pi G(\omega^2\phi^2-V).
\eeq
We find that the kinetic term $\omega^2\phi^2$ , which is responsible for Q-charge, is dominant but vanishes in the center.
Therefore, we call the solutions Q-crusts.

\section{Properties of the solutions}

Stability of Q-balls has been discussed essentially by energetics.
The total energy of the system is defined by
\beq\label{E}
E=\int d^3x\left\{\frac12\omega^2\phi^2+\frac12(\pa_i\phi)^2+V\right\},
\eeq
where $\pa_i$ denotes a spatial derivative.
For a fixed model $V(\phi)$ and the phase assumption (\ref{qball-phase}), there remains a free parameter, $\omega$ or $Q$; accordingly, there is a family of equilibrium solutions.
For such a family, if $E$ increases as a function of $Q$ but $dE/dQ$ decreases, energetics prohibits one Q-ball splitting into two under fixed $Q$. In this case we can understand that Q-balls are stable under the assumption (\ref{qball-phase}).
In connection with this argument, Paccetti Correia and Schmidt showdeed a useful theorem that stability is determined by the sign of $(\omega/Q)dQ/d\omega$ \cite{PCS01}.

Sakai and Sasaki \cite{SS} proposed a simple method of analyzing stability using catastrophe theory \cite{PS78} as follows.
Catastrophe theory reveals stability of a mechanical system completely once {\it behavior variable(s)}, {\it control parameter(s)} and a {\it potential} are given.
Therefore, an essential point is to choose  those variables in the Q-ball system appropriately.
For a given potential $V(\phi)$ and charge $Q$, we consider virtual displacement $\delta\phi(r)$ near the equilibrium solution $\phi(r)$.
If we redefine $\omega$ by
\beq\label{omega}
\omega\equiv Q\Big/\int\phi_{\omega}^2(x)d^3x,
\eeq
the domain of definition of $\omega$ is extended to off-equilibrium configurations.
Using this $\omega$, we can represent a continuous deformation by a one-parameter family of displacement functions, $\delta\phi_{\omega}(r)$.
Then the energy (\ref{E}) is regarded as a function of $\omega$: $E(\omega)\equiv E[\phi_{\omega}]$.
Because $dE/d\omega= (\delta E/\delta\phi_{\omega})d\phi_{\omega}/d\omega= 0$ when $\phi_{\omega}$ is an equilibrium solution, $\omega$ may be regarded as a {\it behavior variable} and $E$ as the {\it potential}. 
The charge $Q$ is given by hand, or physically, it is determined by initial conditions; 
therefore, it should be regarded as a {\it control parameter}. 
Catastrophe theory tells us that stability changes at $dQ/dE=0$ or $dQ/d\omega=0$, which are consistent with the above arguments.

Properties of Q-balls in the model (\ref{V3}) were elucidated as follows \cite{SS}.
\begin{itemize}
\item $\tm^2>1/2$: $V(0)$ is the absolute minimum. There is no bounds on $Q$, and all equilibrium solutions are stable.
\item $\tm^2<1/2$: $V(0)$ is a local minimum but the absolute minimum is located at $\phi\ne0$.
For each $\tm^2$,  there is a maximum charge, $Q_{{\rm max}}$, above which equilibrium solutions do not exist. For $Q <Q_{{\rm max}}$, stable and unstable solutions coexists. 
\end{itemize}
It turns out that properties of Q-tubes and Q-crusts also depend on whether $\tm^2>1/2$ or $\tm^2<1/2$.
Therefore, in the following, we show numerical results for $\tm^2=0.6$ and $\tm^2=0.3$ as typical examples.

\begin{figure}[htbp]
\psfig{file=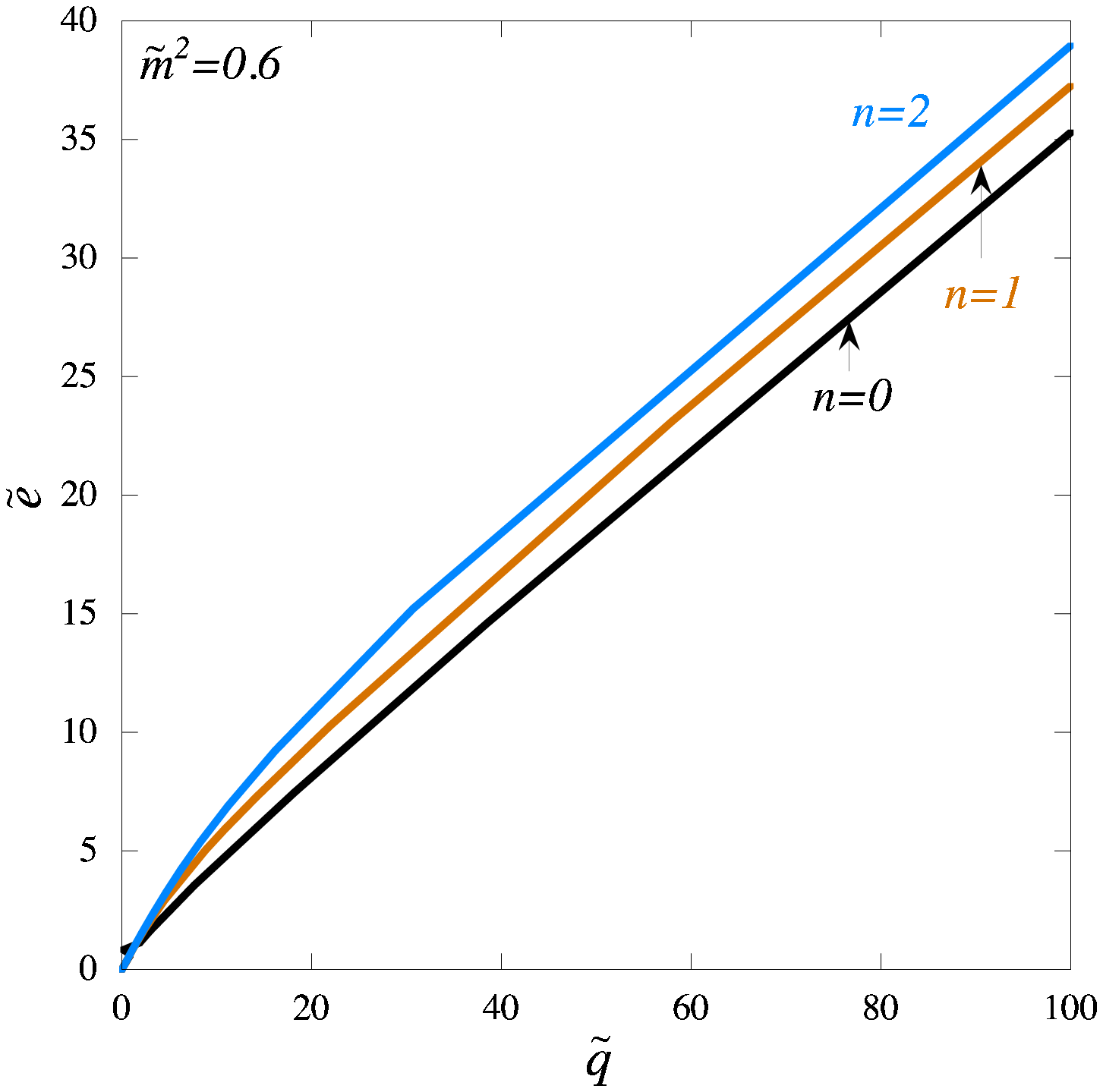,width=3in}
(a)
\psfig{file=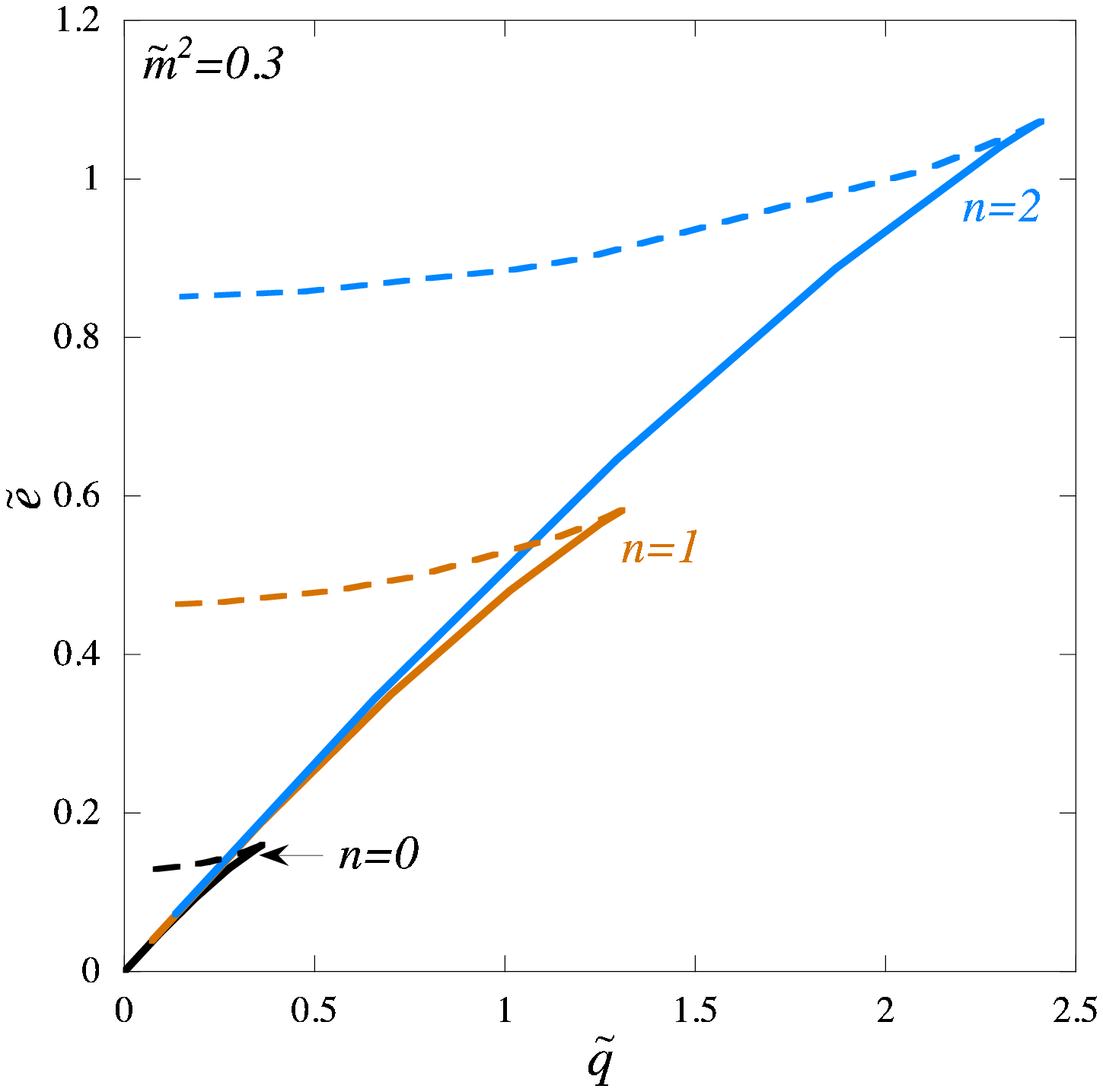,width=3in}
(b)
\caption{\label{QEqtube}
Existence domain of Q-tube solutions in $\tq$-$\te$ space and their stability.
We put $\tm^2=0.6$ and $\tm^2=0.3$  in (a) and in (b), respectively.
The solid and dashed lines represent stable and unstable solutions, respectively. 
}
\end{figure}

\vskip 3mm
{\it (1) Q-tubes.}
In this case, $E$ and $Q$ diverge because they are infinitely long. Therefore, we define the energy and charge per unit length, respectively, as
\bea
{e}&=&2\pi\int_0^{\infty}dR
\left\{{1\over2}\omega^2phi^2+\frac12\left({d\phi\over dR}\right)^2+\frac{n^2\phi^2}{2R^2}+V\right\},\nn
{q}&=&2\pi\omega\int_0^{\infty}\phi^2dR.
\eea
In accordance with the normalization (\ref{rescale}), we rescale the energy/charge variables as
\beq
\te\equiv{\lambda^2\over\mu^2},~~~
\tq\equiv{\lambda^{\frac32}\over\mu}q,~~~
\tE\equiv{\lambda^{\frac32}\over\mu}E,~~~
\tQ\equiv\lambda Q.
\eeq

We show the $\tq$-$\te$ relations for $\tm^2=0.6$ and for $\tm^2=0.3$ in Fig.\ \ref{QEqtube}.
From a viewpoint of these relations, basic properties for Q-tubes are the same as those for Q-balls as described above.
They are summarized as follows.
\begin{itemize}
\item $\tm^2>1/2$: There is no bounds on $q$, and all equilibrium solutions are stable.
For fixed $q$, Q-tubes with lower $n$ are energetically more stable.
\item $\tm^2<1/2$: For each $\tm^2$,  there is a maximum charge, $q_{{\rm max}}$.
For $q <q_{{\rm max}}$, stable and unstable solutions coexists for fixed $q$.
Interestingly, Q-tubes with higher $n$ can have larger $q$.
\end{itemize}

\begin{figure}[htbp]
\psfig{file=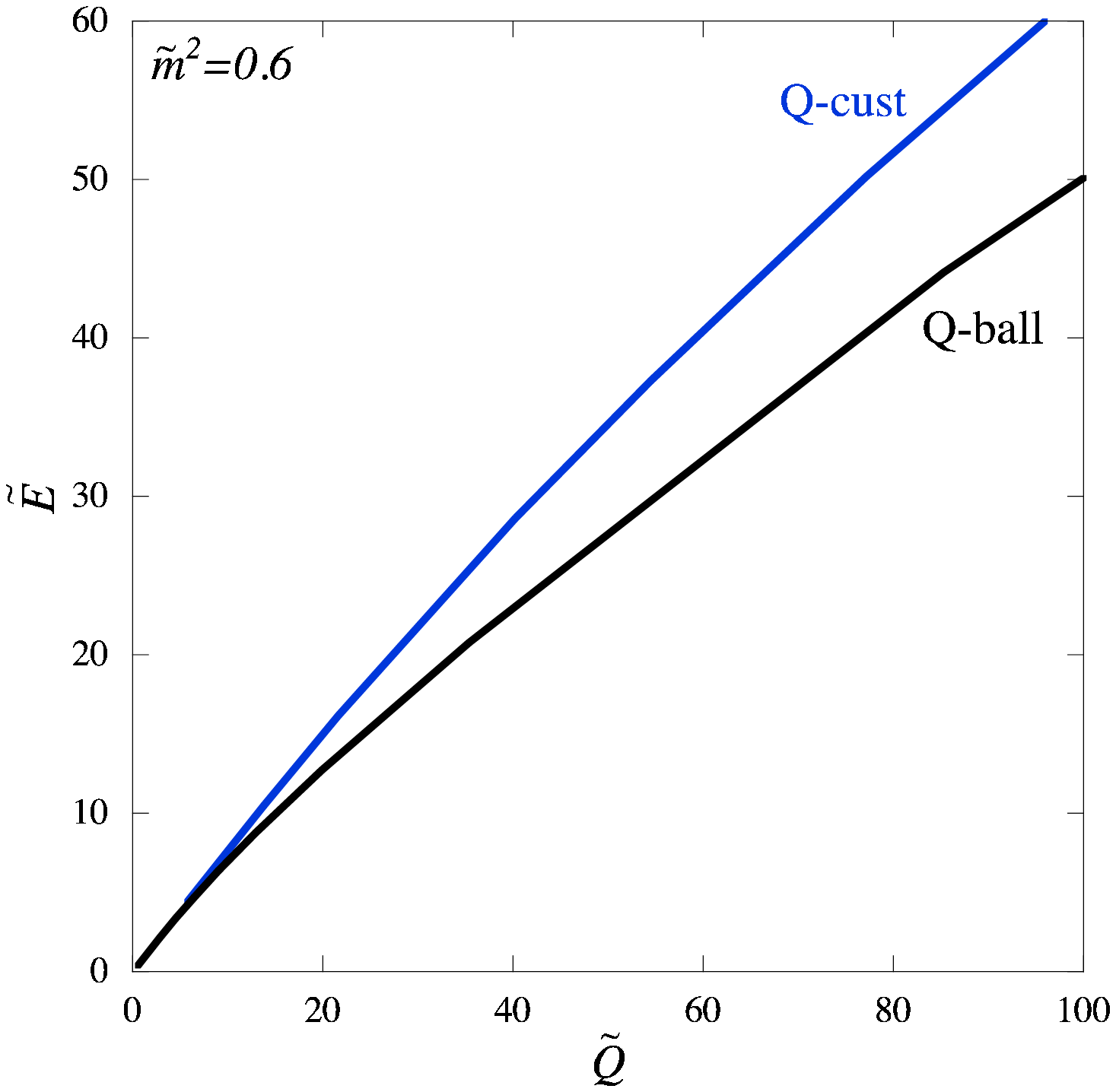,width=3in}
(a)
\psfig{file=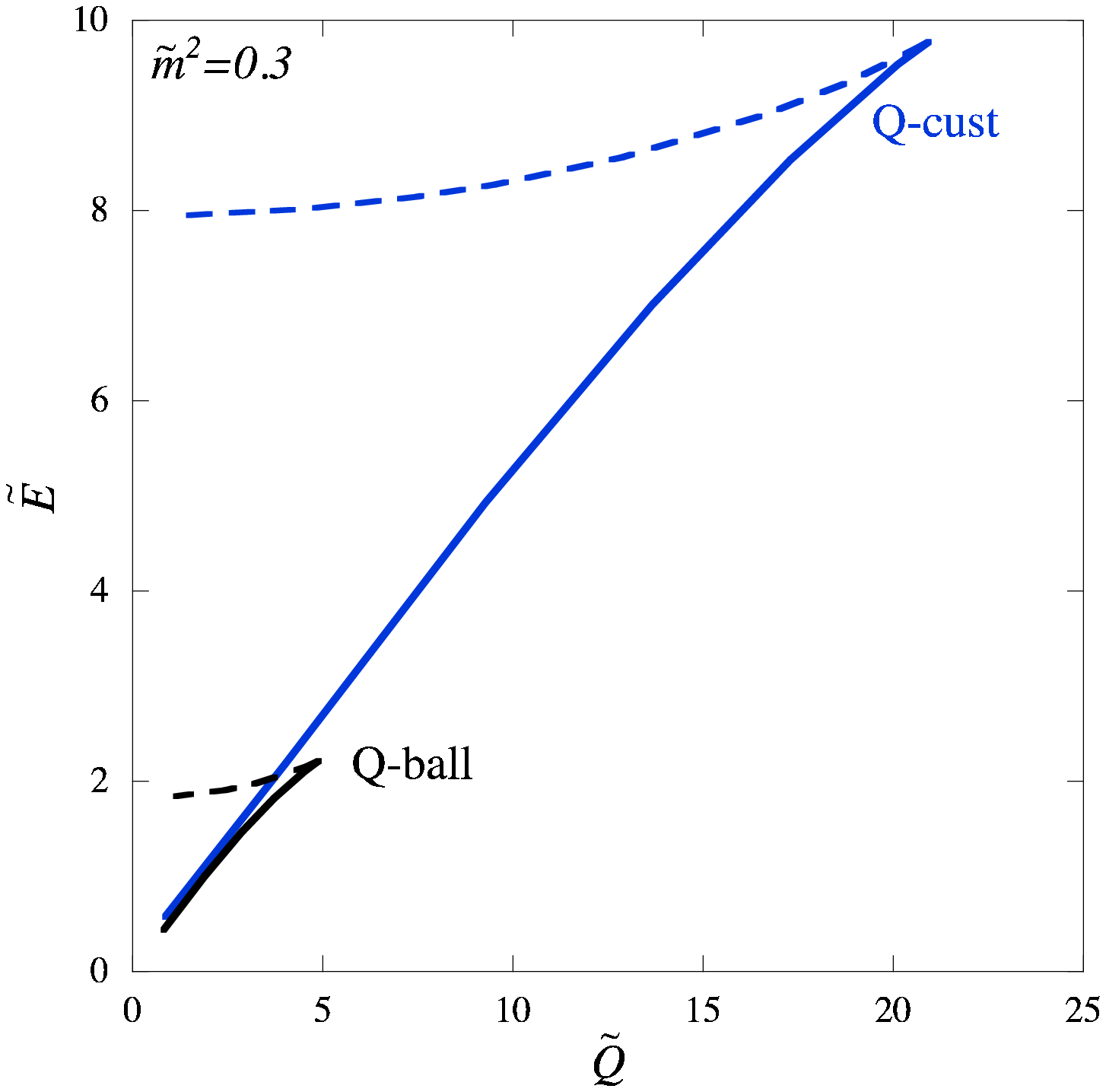,width=3in}
(b)
\caption{\label{QEqcrust}
Existence domain of Q-crust solutions in $\tQ$-$\tE$ space and their stability.
We put $\tm^2=0.6$ and $\tm^2=0.3$  in (a) and in (b), respectively.
}
\end{figure}

\vskip 3mm
{\it (2) Q-crusts.}
Similarly, we show the $\tQ$-$\tE$ relations for $\tm^2=0.6$ and for $\tm^2=0.3$ in Fig.\ \ref{QEqcrust}.
For reference, we show the results for Q-balls, too.
(Note that Q-balls can appear not only in U(1) theories but also in a wide class of theories which include global U(1) symmetry.)
What we find for Q-crusts is summarized as follows.
\begin{itemize}
\item $\tm^2>1/2$: There is no bounds on $Q$, and all equilibrium solutions are stable.
For fixed $Q$, Q-balls are energetically more stable than Q-crusts, as expected.
\item $\tm^2<1/2$: For each $\tm^2$,  there is a maximum charge, $Q_{{\rm max}}$.
For $Q <Q_{{\rm max}}$, stable and unstable solutions coexists.
Interestingly, Q-crusts can have larger $Q$ than Q-balls.
\end{itemize}

\vskip 3mm
We should note that the above results are based on the phase assumption (\ref{qtube-phase}) or (\ref{qcrust-phase}).
Stability against perturbations on the phase configuration (\ref{qtube-phase}) or (\ref{qcrust-phase}) cannot be revealed by the present energetic analysis and should be studied by dynamical analysis.

\section{Summary and Discussions}

We explore equilibrium solutions of non-topological solitons in a general class of scalar field theories which include global U(1) symmetry, and then we find new types of solitons:
Q-tubes in U(1) theories and Q-crusts in SO(3)$\times$U(1) theories.
Because only $n=0$ Q-tubes have homogeneous phase like Q-balls, they can appear in a wide class of theories which include U (1) symmetry. 
Except for $n=0$ Q-tubes, there is a dip in kinetic energy in the center. 
In contrast with cosmic global strings or global monopoles, their gravitational mass can be finite without gauge fields.

We also investigate stability of equilibrium solutions for the model (\ref{V3}) under the phase assumption (\ref{qtube-phase}) or  (\ref{qcrust-phase}) by calculating their charge and energy (or those per unit length).
The charge-energy relations indicate that, if $V(0)$ is the absolute minimum, there is no bound on charge,
and all solutions are stable.
On the other hand, if $V(0)$ is a local minimum but the absolute minimum is located at $\phi\ne0$, there is a maximum charge, above which equilibrium solutions do not exist. 
For fixed charge below the maximum, stable and unstable solutions coexists .
It is interesting that Q-tubes with higher winding number can have larger charge density and that Q-crusts can have larger charger charge than Q-balls.
Stability against these phase configurations is beyond the present energetic analysis and should be studied by dynamical analysis, which is our next subject.

Unlike Q-crusts, our Q-tubes solutions are infinitely-long and unrealistic in themselves.
Nevertheless, Q-tubes are the more interesting because they can appear in the minimal (i.e., U(1)) models and several researchers have already performed numerical simulations in those models \cite{EJ01}.
Those simulations showed that filament structure appears just before Q-ball formation and maintain its shape for a certain time. 
We conjecture that such a filament structure is semi Q-tubes.
Furthermore, according to recent simulations of the collision of two Q-balls \cite{Tsuma}, apparent two rings are formed.
We suspect that they are loop Q-tubes.
Further fundamental investigations of Q-tubes and the other solitons together with advanced simulations of Q-ball formation may confirm the above conjectures and elucidate observational consequences of Q-ball formation in SUSY. It may be also interesting to explore analogous new solitons in non-relativistic atomic Bose-Einstein condensates \cite{EL}.

\acknowledgments
We thank T. Hiramatsu and T. Tamaki for useful comments.
The substantial part of this work was done while NS stayed at Osaka City University. 
NS thanks his colleagues there for their hospitality.
This work was supported by Grant-in-Aid for Scientific Research Nos.\ 22111502, 19540305 and 21540276.

\end{document}